# A Framework for Operations Research Model Use in Resilience to Fundamental Surprise Events: Observations from University Operations during COVID-19

Thomas C. Sharkey, Steven Foster, Sudeep Hegde, Mary E. Kurz, and Emily L. Tucker

Department of Industrial Engineering, Clemson University

Corresponding author: Thomas C. Sharkey, tcshark@clemson.edu

**Abstract**

Operations research (OR) approaches have been increasingly applied to model the resilience of a system to surprise events. In order to model a surprise event, one must have an understanding of its characteristics, which then become parameters, decisions, and/or constraints in the resulting model. This means that these models cannot (directly) handle fundamental surprise events, which are events that could not be defined before they happen. However, OR models may be adapted, improvised, or created during a fundamental surprise event, such as the COVID-19 pandemic, to help respond to it. We provide a framework for how OR models were applied by a university in response to the pandemic, thus helping to understand the role of OR models during fundamental surprise events. Our framework includes the following adaptations: adapting data, adding constraints, model switching, pulling from the modeling toolkit, and creating a new model. Each of these adaptations is formally presented, with supporting evidence gathered through interviews with modelers and users involved in the university response to the pandemic. We discuss the implications of this framework for both OR and resilience.

# 1 Introduction

Operations research (OR) models, and other analytics, have been increasingly created to analyze the resilience of a variety of systems, including civil infrastructure and supply chains. Common themes in the definitions of resilience focus on the ability of a system to respond, adapt, or recover when faced with stressful events, based on the foundational work of Holling (1973). In order to *model* a *stressful event*, one must have an understanding of its characteristics which become inputs into the OR model. Therefore, by definition, stressful events that appear in OR models for resilience must fall into the category of *situational surprise* as discussed by Lanir (1986) and Wears and Webb (2014). Alternatively, *fundamental surprise* involves events that "*refute basic beliefs about 'how things work'*" (Wears and Webb, 2014) which implies that they fundamentally alter the environment in which the system is operating; fundamental surprise events arise from issues where one "*cannot define in advance the issues for which one must be alert*" (Wears and Webb, 2014). Therefore, any stressful event that is considered by an OR model prior to the event must be a situational surprise event since the characteristics of the event must be defined a priori.

This does not mean that OR models cannot play a role in the response to fundamental surprise events but suggests that they must be deployed *during* or *in response to* the event. This leads to the following question: How can OR models be used to improve resilience to fundamental surprise events? Given the limited number of fundamental surprise events that occur, gathering data to answer this question is challenging. Further, fundamental surprise events necessitate improvisation in many aspects, including how analytics are deployed. Eisenberg et al. (2019) discuss that improvisation by the modeler, the user, or the model of "resilience analytics" is necessary to deploy analytics in response to fundamental surprise. They present examples of this improvisation for fundamental surprise events including Hurricane Ophelia that struck Europe in 2017 (Mersereau, 2017), the near-breaching of the Oroville Dam in 2017 (Hollins et al., 2018), and the Arizona-Southern California power outage of 2011 (Clark et al., 2018).

Woods (2015) provides four ways in which systems can demonstrate their resilience which are rebound, robustness, graceful extensibility, and adaptability. Resilience as rebound focuses on the ability of a system to recover from a stressful event. Resilience as robustness focuses on the ability of the system to maintain operations under stressful events. Graceful extensibility focuses on the ability of the system to alter boundaries or operations when surprise events challenge it. Adaptability focuses on the ability of the system to adapt to future stressful events as conditions evolve. When a system is designed or decisions are made with an OR model, the potential for the system or decisions to be resilient to events that were not considered by the model may be severely limited since extensibility or adaptability considerations rely on using the system in unforeseen ways.

Sharkey et al. (2021) framed the contributions of the network optimization community, a subset of the broader OR community, to resilience within the four concepts of Woods (2015). These contributions tend to focus on resilience as rebound (Çelik 2016; Sharkey et al., 2021) and resilience as robustness (see, e.g., Snyder, 2006; Çelik et al., 2012; Faturechi and Miller-Hooks, 2015; Snyder et al., 2015; Wang et al., 2016; Hosseinni et al., 2019), consistent with the differences between situational and fundamental surprise events. Extensibility and adaptability often require improvisation either within the model or by those creating/using the model during a fundamental surprise event.

Eisenberg et al. (2019) provide a framing mechanism to understand how improvisation would occur by the model, the user of the model, or the modeler (or some combination of them) to address a fundamental surprise event, increasing the extensibility and adaptability of models. Sharkey et al. (2021) highlights that optimization models that enable improvisation are an important research direction.  To move towards creating models that enable improvisation, we should first seek to understand how models are deployed in fundamental surprise events.  The focus of this paper is to provide a classification of how OR models were deployed in the response to COVID-19 at Clemson University.  COVID-19 was a fundamental surprise event since it altered the basic structure of how courses were delivered (e.g., in-person instruction had to be cancelled and/or done with social distancing in mind), how resources (such as hand sanitizer) needed to be deployed, and how distinct organizations within the university needed to coordinate efforts.  We provide empirical evidence of the OR modeling improvisations and adaptations that took place during the response to COVID-19 and a classification of the types of adaptations that were used in practice. This empirical evidence could help to identify modeling guidelines so that models can contribute to graceful extensibility and adaptability.

The process to improvise OR models during fundamental surprise events is like the general OR modeling process.  Morris (1967) discusses the types of tasks that are necessary to inform mathematical models.  Willemain (1994, 1995) interviewed numerous OR experts to understand how they create models when faced with a practical problem.  This creation process was especially important given the number of stakeholders that experts or, using the terminology of Eisenberg et al. (2019), 'modelers,' had to interact with and consider while creating models, conducting analysis, and communicating their recommendations.  Pidd (1999) offers six principles for modeling in practice, with several of them being relevant to our classification scheme. The principle of model development being a gradual process is illustrated in our adaptations where models are either pulled from a toolkit or created to handle a unique feature within the university response. The principle of proper use of data is important since the fundamental surprise event radically changes the operating conditions of models that were deployed prior to the event. The principle of the modeling process being chaotic certainly relates to the time-critical nature of some of the decisions that were being made with newly deployed models. The contextual considerations during model creation of Gorman (2021) emphasize the need for modelers and users to work carefully along with stakeholders as they try to use models to support decision-making, which is a consideration of OR model use in response to COVID-19.

The response of Clemson University to COVID-19, like all colleges and universities, involved adaptations and new challenges across all its organizations (Hegde et al., 2022). Clemson University moved all of its Spring 2020 classes online on March 19, 2020 and announced that its Summer 2020 classes would be online on April 8, 2020. It was committed to providing in-person classes in Fall 2020, while adhering to federal COVID-19 guidelines for social distancing and other public health measures. Three areas in which OR approaches were considered were the scheduling of classrooms, scheduling of cohorts, and locating hand sanitizing stations on campus.

Table 1 provides an overview of the pre-pandemic approaches to the problems of classroom scheduling, cohort scheduling for hybrid classes, and hand sanitizer location as well as how (and when) changes to the modeling approaches occurred. Classroom capacities were a significant factor in classroom scheduling. Cohort scheduling, which hoped to identify sets of students to all attend in-person class on the same day, became important during hybrid teaching where students would be assigned to a single weekly "in-person" day per class. Hand sanitizer location and distribution was needed to help follow public health guidelines as faculty, students, and staff interacted in-person.

| | Pre-pandemic | Fall 2020 | Spring 2021 |
|---|---|---|---|
| Classroom Scheduling | Banner, with historical classroom data, and input from department-level schedulers and students | Banner, with classroom data update in accordance with COVID-19 guidelines | Banner, with classroom data update in accordance with COVID-19 guidelines |
| Cohort Scheduling (Hybrid) | Policy not in place: No model needed | Cohort-based MIP model | Cohort-based MIP model |
| Hand Sanitizer Location | Resource not deployed: No model needed | Heuristic-driven demand model. (No math modeling) | Application of facility location models |

*Table 1: An Overview of Model Use and Adaptations by the COVID-19 Response of Clemson University*

Clemson University's normal class scheduling process works as follows. Classes are typically either Monday/Wednesday/Friday (MWF) or Tuesday/Thursday (TTh), with common variants being once-a-week courses (labs) or four-times-a-week courses (calculus courses). Each day has standard course times (i.e., 8 am – 8:50 for MWF or 8 am – 9:15 am for TTh courses). Each department has a course scheduler who determines when each course should be delivered and estimates the size of each course. A very small proportion of classes can be assigned to "department" rooms, which are not centrally scheduled, while the majority (>95%) will be scheduled in "university" rooms. The university registrar then allocates classrooms to courses.

In preparing for the Fall 2020 semester, social distancing policies impacted classroom capacity. Based further on the desire to mitigate the spread of COVID-19 while allowing as much in-person instruction as possible, instructors had the option to request fully online courses or in-person courses. Students could elect to be fully online. Once the initial set of selections was analyzed, there was not enough classroom capacity to allow traditional in-person instruction. As described in Gore et al. (2022), Clemson University enlisted a team of faculty and graduate students to create a rotational attendance scheme in which each student is assigned to exactly one "in-person" day a week in each course, resulting in hybrid courses. This team created an integer program (IP) to determine the rotational attendance scheme. On July 22, 2020, the university announced that the first four weeks of the Fall 2020 semester would be online, providing more time to plan for hybrid courses. When course planning for Spring 2021 began, knowing that social distancing would be required, departmental administrators could determine course modalities, including online, traditional (all in-person) or hybrid (rotational attendance). Course timings and capacities were set and a centralized room scheduler was run, after which departments adjusted schedules to facilitate hybrid or traditional modes. Students were able to register, knowing the course schedules and teaching modes as impacted by social distancing in the classrooms and the same IP was applied.

The Facilities Department at Clemson University was responsible for deploying hand sanitizer across campus to mitigate the spread of COVID-19. They purchased hand sanitizer and built sanitizer stations that could be placed at entrances to different buildings. For Fall 2020, they determined locations for these stations using heuristics and building coordinator familiarity with relative door usage. In Spring 2021, OR modelers worked with the department to develop facility location models to recommend revised locations based on estimated demand from quantitative door access data. O'Brien et al. (2022) discuss a blended approach that integrates facility location modeling with rigorous qualitative interview analysis that was applied to reach these recommended solutions.

In general, the delivery of academic programs is the responsibility of college or university staff. As summarized in Barnhart et al. (2022), many aspects of delivering academic programs have been addressed in the literature. These aspects include term planning, time tabling and room assignment, and student enrollment. As institutions faced the disruptions of the COVID-19 pandemic, especially entering the 2020-2021 academic year, these aspects were altered in unexpected ways (see, for example, University of Pittsburgh Center for Teaching and Learning, 2021 and Cornell University Office of the University Registrar, 2021). Analyzing the response to COVID-19 at universities offers a unique perspective on how this fundamental surprise event impacted OR model use.

OR models were created for a variety of decision-making problems in response to the fundamental surprise event of COVID-19, outside of universities. Shen (2020) provides an overview of how optimization models can support COVID-19 response in a variety of manners. Currie et al. (2020) provide a similar overview for simulation models. Mehrotra et al. (2020) examined resource sharing of ventilators and applied it to analyzing their FEMA stockpile during COVID-19. Bastani et al. (2021a, 2021b) describe how operations research and reinforcement learning can support deploying limited testing resources as people move across international borders to help prevent the spread of COVID-19 and applied them to border crossings into Greece. Given the broad scope of COVID-19 OR modeling, we will focus on a subset of this work (university operations) in the context of our empirical data collection.

The contributions of this paper are focused around providing a better understanding of how OR models were used in response to the fundamental surprise event of COVID-19 through the case study of a university. We provide a classification scheme of uses in terms of how models were adapted, switched, or created. This classification scheme is based on interviews with professionals involved in the decision-making processes for each of the three areas discussed in Table 1. We provide evidence and examples from these interviews for each of our classifications, which helps to provide more qualitative support for the improvisation schemes laid out by Eisenberg et al. (2019). We review a sampling of other OR models deployed at universities in their response to COVID-19 and apply our classification framework to them. We also discuss potential research directions at the intersection of human factors methods and OR modeling with a focus on supporting organizational resilience.

The remainder of this paper is organized as follows. We provide our research methodologies, classification scheme, and examples in Section 2. Using our classification scheme, we analyze a sampling of OR work done at other universities in Section 3. We discuss additional insights in Section 4 and lay out an agenda to better enable the use of OR models during fundamental surprise events. We conclude in Section 5.

## 2 Classification Scheme of Model Use during Fundamental Surprise Events and Supporting Evidence

The focus of this section is on the classification scheme of how OR models may be used to support the response to a fundamental surprise event. We use the response to COVID-19 on a university campus as a case study for gathering our empirical evidence. Following the framework of Eisenberg et al. (2019), we examine how adaptations or improvisations were necessary by the *model*, *modeler*, or *user*. The model focuses on the underlying mathematical representation of the decision-making environment. The modeler is the person (or people) responsible for creating the mathematical model. The user is the person (or people) that will run the model and recommend decisions based on its output. We begin the section with an overview of our research methodologies to gather the evidence and then provide an overview of the framework. We then take a deeper look at each area in the classification scheme, along with supporting evidence.

### 2.1 Research Methods

This study was conducted using qualitative analysis of semi-structured interviews. Participants were comprised of 14 pandemic modelers and users from a large public university in the southeastern United States. These participants helped to coordinate the university's pandemic response, specifically with regards to class scheduling and resource allocation. Participants include administrators and coordinators from the university Registrar's office as well as department-level scheduling coordinators and analysts.

The semi-structured interview script (Appendix) was developed to elicit considerations for class scheduling and resource allocation with the goal of understanding the university's strategies for organizational adaptation in response to the COVID-19 pandemic. Questions were designed to identify participants' roles in relation to the university's pandemic response and identify specific constraints and objectives that ultimately shaped the decisions and policies of the university. Further probes were based on prior knowledge elicitation and used to expand upon aspects of adaptation such as improvised workflows, resource constraints, monitoring capabilities, and coordination between departments, stakeholders, and other groups on campus.

Analysis of interview data was conducted using qualitative content analysis carried out through MaxQDA 2022 (VERBI Software, 2021). Themes were identified from the interview data and connected with traditional mathematical modeling considerations to better understand ways in which large organizations can use OR to adapt to fundamental surprise events. After gathering data on modeling considerations, the OR experts on this paper reviewed the transcripts to identify potential passages that would be relevant to modeling efforts.

Another important note is that Clemson University was not using an OR model for course scheduling. However, it would not be difficult to build a new class scheduling and classroom allocation optimization model or apply an existing one. Therefore, we view the class scheduling process as the underlying model and believe that we can draw conclusions about how adaptations to this process would translate to using OR models during fundamental surprise events.

## 2.2 Overview of Classification Scheme

We identified five different areas for OR model use during the response to this fundamental surprise event. Overall, we found that OR modeling use during periods of fundamental surprise involves model adaptation, switching, and creation. Based on interviews and discussions, we classify the following model use in five areas: adapt data, add constraints; switch model; pull model from toolkit; and create new model. Table 2 provides an overview of each of these classifications, a summary of the model characteristics, and who or what is making the adaptation. For each classification, we provide their definition, an overview and examples from our data, and practical considerations for their use.

| Category | Definition | Model Characteristics vs. Pre-Surprise | | | Adaptation by |
|---|---|---|---|---|---|
| | | Decision | Model | Data | |
| Adapt Data | The underlying model remains the same but old data is no longer useful | Same | Same | New | User |
| Add Constraints | The underlying model remains the same, but the decision-maker gives less freedom to the model. | Same | Modified | New | Either modeler or user |
| Model Switch | The underlying model is substantially revised or switches to a different model altogether | Same | New or modified existing | New | Modeler and user |
| Pull Model from Toolkit | A new decision that needs to be made due to the surprise event could be made better by applying existing modeling approaches. | New | New | New | Modeler and user |
| Create New Model | An entirely new model is needed for effective decision-making. | New | New | New | Modeler and user |

*Table 2: An overview of our classification scheme for how OR models are used in response to fundamental surprise events*

**2.3 Adapting the Data**

**Definition:** The underlying model remains the same but traditional (or historic) data is no longer useful and, therefore, data needs to be adapted to represent the current situation.

**Overview and Examples:** Adapting the data occurs when the user believes that the fundamental surprise event does not fundamentally break the assumptions of the model but does break traditional data streams that go into it. This can be viewed as an adaptation mainly by the user since there is no need to update the model itself. Many examples arose in COVID-19, especially around planning for the return to in-person classes while accounting for social distancing. Traditional data on classroom capacities was fundamentally inaccurate when considering requirements for students to be socially distant from one another. Further, data on which on-campus facilities and rooms could be used to hold classes also was no longer valid. For example, at Clemson University, basketball courts in the recreational gym were used as classrooms to increase the size of courses that could be taught in-person under social distancing guidelines.

Data on classroom capacities, and how to accurately and quickly capture it, was important in the class scheduling process. In fact, the following quotes illustrate how modeling used estimates of new data:

> *"At that point we didn't have room capacities for our initial modeling. For every room, we extracted out the largest class that was registered in that room and then divided the capacity by a third just for our initial testing."*
>
> *"They said to use a six-foot rule and so from there we calculated this is how many people were in the class."*

As the decision-making process evolved, better methods to estimate data for certain classroom types were implemented:

> *"It was determined quickly that for square classrooms, you could, with movable seating, you could sort of do square footage divided by 36 plus one for the instructor and get a good ballpark estimate.*

Those gathering the adapted data may need varied methods, even for similar parameters within the model (e.g, classroom capacities).

Data on the set of rooms available to schedule classes in (the resources in a classroom scheduling model) required inputs from multiple organizations:

> *"It was a fairly large team, it was 10ish people, and we evaluated data about the spaces and whether or not a room is suitable for us to use as a classroom, especially like the big box spaces and determining whether we could hold a class in, say, a gymnasium. … [There was] a group from facilities, including the space management group, facilities project manager, some of the people from campus planning and university planning and design …. and then, of course, we had the registrar's group that fully understood the schedule and related nuances."*

> *"We came up with the idea of using five gyms as our entry-level 101 classes, and so we spent a bunch of money and worked with the people over in [the campus rec gym] to make sure they could still try to safely operate while holding classes."*

Adapting the data into the model may not always fall to a single person since the improvisation needed to understand how resources can be used may need to be made by various stakeholders with different knowledge bases.

**Practical Considerations for Adaptation Use:** This adaptation requires that an underlying model currently exists and is being deployed to support decisions. It requires that users and decision-makers understand the underlying assumptions of the model and have enough information about the fundamental surprise event to believe the model is still useful if the right data is gathered.

The *data-gathering process* of the organization(s) impacts this adaptation because it must be appropriately accurate for the decision-making horizon. In the example of classroom scheduling, this data-gathering process included: (i) updating capacities of traditional classrooms and lecture halls (model parameters), and (ii) updating the schedulable spaces (the sets and decision variables). For Fall 2020, Clemson University had close to four months to gather such data. If this time horizon was smaller, then it may not have been possible for this to occur.

The intersection of the data-gathering process and the adaptability of the decisions also plays a role in the potential use of this adaptation. If the data-gathering process may take longer than when the first decisions must be made, then this adaptation can still be useful if the decisions can be updated. The classroom scheduling example is one where the decisions are not easily adapted since it was a highly constrained environment in that moving one class could have significant ripple effects. Therefore, data needed to be gathered prior to the deployment of the decisions.

## 2.4 Adding Constraints

**Definition:** The underlying model remains the same, but the modeler or the user provides it less freedom to help address issues around lack of information and/or uncertainty.

**Overview and Examples:** This adaptation focuses on the modeler or the user identifying considerations that arose from the fundamental surprise event and adding constraints directly to the underlying model that help capture these considerations. An important special case is when the constraint "fixes" a variable to a specific value. In the context of Eisenberg et al. (2019), this adaptation may be implemented by the user if the way they interact with the model is sophisticated enough to allow for them to add new constraints or it may require an interaction between the modeler and the user. For example, a user may be able to directly add a constraint that fixes a variable; a constraint that focuses on a set of variables may need to be incorporated into the model by the modeler.

For course scheduling, an important consideration was to determine which classes would be fully online and, therefore, would not need to request a classroom (resource) within the class scheduling model. In this case, the added constraint would be to enforce that the class is online or, equivalently, all variables associated with assigning a classroom to the class would be zero. For example, one department indicated that large classes would have been difficult to accommodate in classrooms in a socially distant manner:

> *"[O]ur largest classes are MSE [Materials Science and Engineering] 2100. The department here said, 'Those are online. We're not going to try because there's no way we can find a room.' The other class we have is materials processing, MSE 3190 that has BIOE [Bioengineering] and MSE students, that's typically 150. Again, 'That's going to be online, we're not going to be in-person.' Our other class sizes were much more reasonably sized with anywhere from 15 to 40-45 students at most. So that was why we had no problem getting a room."*

There were further issues surrounding the modalities by which classes were offered, especially when considering that most campus buildings were not allowed to be accessed in a regular fashion. For example, the library required reservations for students to enter which made it difficult for students that may have been going from an in-person class to an online class with only 15 minutes between them:

> *"The biggest thing for my students was if they had classes back-to-back, and they were going to have to, you know, a lot of times they, and, if they really wanted to be in person, it was going to be too difficult. So they had to find a place on campus and just attend online. And then I had students that really, really needed to be in person all the time."*

Constraints could be added that appropriately spaced-out classes within a particular curriculum (e.g., courses typically taken together during fall semester of a particular year) that had different modalities. However, as will be discussed later, the actual modality that was implemented by a particular instructor may have been adapted on the fly which means this situation may have arisen for unexpected course pairings.

**Practical Considerations for Adaptation Use:** This adaptation requires that an underlying model currently exists and is being deployed to support decisions. It requires that users understand the underlying assumptions of the model and have enough information about the fundamental surprise event to understand the considerations that need to be updated to make the decisions the model is capturing during the event.

A successful application of this adaptation requires differentiation between *hard* constraints and *soft* constraints. This is especially important as the user surveys the situation after the fundamental surprise event. The "conventional wisdom" was that all classes should be offered in-person as much as possible, and that some in-person teaching for each class was required. When some departments realized that a fully-online experience could provide a more uniform teaching and learning experience, ***and*** reduce demand for large classrooms, other classes were more able to find appropriately sized classrooms. In this way, a potential hard constraint that required every class to have some in-person component was removed, and a subset of classes had their "offering" variable set to "online only" and their related classroom assignment variable was fixed.

**2.5 Model Switching**

**Definition:** There was an underlying model in use, but the surprise event requires a *switch* to a different (existing) model.

**Overview and Examples:** This adaptation focuses on examining the current environment after the surprise event, realizing that the current model no longer applies, and switching to an existing model that better applies to the post-event environment. That is, the decision to be made is similar, but the context surrounding the decision is not. Therefore, it involves both adaptations by the user in recognizing that the current model is no longer sufficient and by the modeler in working with the user to adapt to a better model for the current situation. Specific adaptations may include changing objectives to constraints (like an epsilon approach to multi-objective optimization) or changing objectives completely.

Sharkey et al. (2021) describe model switching where a network under normal operating conditions seeks to meet demand at minimum cost but once a disruption occurs it switches to meeting as much demand as possible. Once repairs to the network are sufficient to meet all demand, there is another switch where the network goes back to meeting demand at minimum cost. This illustrates important points for model switching including: (1) often what was a constraint in the original model (e.g., meeting demand) becomes the objective in the adopted model and (2) constraints in the original model (e.g., meeting demand) may no longer be feasible after the surprise event, thus rendering the original model irrelevant.

During the COVID-19 response, models were switched for deploying hand sanitizer. Initial decisions were made to best deploy limited resources (e.g., hand sanitizer 'stands') but eventually campus became saturated with hand sanitizer:

> *"Of course, everything was eventually saturated but at the start we had very limited supply, so we built some stands to be able to put those hand sanitizers out in our primary use buildings."*

The initial objective was ensuring that all people on campus would be "covered" by the deployed stands:

> *"From my understanding, we were trying to cover everybody that was on campus."*

Once everybody was covered, campus wanted to minimize the costs of deploying hand sanitizer (new objective) while ensuring the coverage of hand sanitizer (old objective, now a constraint).

Another example of model switching is the incorporation of safety considerations into decision-making surrounding campus activities:

> *"I think that maybe the EOC [Emergency Operations Center] at times will make sure the students are safe and make sure that the people that were provided that material, our custodial staff, had all their PPE [Personal Protective Equipment] stuff and they had their mask and their personal sanitizing wipes so they would try to make sure those guys were safe."*

Although general safety considerations are incorporated into campus maintenance, safety considerations surrounding disease spread were not at the forefront of these decisions prior to COVID-19.

**Practical Considerations for Adaptation Use:** This adaptation not only requires that an underlying model is in use but there is another model that can be easily switched to after a decision-maker realizes the limitations of the existing underlying model. It requires the user to identify the limitations, including when certain constraints could no longer be satisfied and/or which would become objectives, and how other existing models would overcome them including when certain constraints

**2.6 Pulling from the Modeling Toolkit**

**Definition:** A new decision needs to be made due to the surprise event and can be made by applying variants of classic or pre-existing optimization models.

**Overview and Examples:** This adaptation focuses on the development of a model to address a new decision brought on by the surprise event. The modeler uses an existing optimization framework, i.e., "pulling from the modeling toolkit," to create a model in partnership with the users and other stakeholders. The classic model may need to be supplemented with minor side constraints, but the core of the approach can be developed within an existing model structure.

At Clemson, a new decision that needed to be made during COVID-19 response was the location of hand sanitizer stations in academic buildings. Prior to the pandemic, hand sanitizing occurred in restrooms or with individual-carried hand sanitizer bottles; there were no hand sanitizer stations. To prepare campus for student return, the Facilities Department was directed to locate sanitizer stations throughout campus:

> *"Hand sanitizer wasn't something that we as a facilities department stock for every building on campus. ... as far as most of the academic areas there really weren't many [with sanitizer] and we quickly realized that was going to be a problem. ... Once we realized what we could get in terms of sanitizer stations, then we had to start making hard decisions on where those stations should be located and where the most impactful locations are for any additional sanitizers we could possibly add."*

The initial allocation occurred quickly and without a model:

> *"[The stations] were initially located primarily via expert judgment ... from [those]... who are responsible for the infrastructure and maintenance of each building on campus."*

To improve allocation, modelers partnered with the Facilities Department to develop an optimization model to recommend locations. Within the proposed classification scheme, this adaptation involves: (i) a new decision (hand sanitizer locations); (ii) a new model to make recommendations on where to locate the stations; and (iii) new data. Because the decision context fit primarily within an existing framework – facility location models – the modelers could pull directly from the toolkit during model development:

> *"The main goal was getting students faculty and staff ... access to hand sanitizer stations. ... What we did on our [OR modeling] team was figure out ways to help allocate sanitizer stations effectively to be able to achieve that"*

During model development, it was important to understand the context – including the goal (objective), decisions to be made, restrictions on the decision, the ability to implement the recommended decisions, and data available. Each facet informed what type of modeling framework would be appropriate. Partnership with users and other stakeholders was critical:

> *"Facilities informed every aspect of the model. From [December 2020] forward, we would meet with them every other week for about a year or so. We had several conversations with them about the structure of the model. We... didn't talk about constraints or objectives, but rather we asked questions and had conversations to get at ... their priorities"*

Several versions of the model were created and discussed with stakeholders:

> *"We ... consider[ed] several modeling structures, but we ended up landing on a facility location structure. ... We selected [this] because of the priorities of the Facilities Department. For example, when we are considering a network flow style approach we had brainstormed about possibly locating hand sanitizer stations outside of the buildings; however, when that came up with facilities, they said no, no, the stations have to be in the interior. I think one reason was because of theft, and that was subject matter expertise that we wouldn't have been able to consider without their input."*

Through each of the iterations, the models were all "based on a conventional modeling structure":

> *"at one point we were considering a p-median based model. Through different model iterations and subsequent conversations with the Facilities Department we ended up landing on a maximum coverage style modeling approach."*

The Facilities Department guided what version of the model was considered final:

> *"There were opportunities to further enhance the model, whether we were going to consider interior spaces within the buildings, as opposed to simply the doors but there came a point where the Facilities Department [were] ... satisfied with the model that we presented. At their behest, we decided to stop iterating, and so it was not only the structure of the model, but rather the decision to complete the modeling effort that came from discussions with the Facilities Department."*

In this example, the final adopted model was a variant of the classic maximum coverage model (O'Brien et al. 2022), in which a set of candidate locations (doors), either have a hand sanitizer station installed or not (binary decision variable) to maximize the number of people who use those doors with hand sanitizer stations installed (objective), subject to a limited hand sanitizer budget.

**Practical Considerations for Adaptation:** A toolkit-style adaptation is useful when the surprise event causes a new decision to need to be made, there is enough time in the build-up to the decision (or the decisions are adaptable after initially made) that a model can be implemented, the decision fits within the context of an optimization model used in another setting, and the users and stakeholders can collaborate with modelers.

Throughout the process, integrated development with stakeholders is critical. It may be worthwhile to consider an iterative approach to model development that incorporates qualitative interviewing of stakeholders (O'Brien et al. 2022). The process of developing model iterations was conducted in a feedback loop with the Facilities Department review and input at many stages. Model development itself can be expedited by using an existing modeling framework.

Modelers should also seek to understand the full range of users and stakeholders and coordinate to the degree possible. There may be multiple stakeholders seeking to address the new decision. With hand sanitizer stations, *"Some of the individual departments or colleges had previously gone out on their own and put in sanitizer stations."* Because the decision is new, there may not be a mechanism for implementation of the model recommendations, and modelers should take care to discuss this process with users and stakeholders.

The context may continue to change after the surprise event. More information may become available, and modelers may be able to incorporate it or adjust the model to address. In this example, the university began to record door accesses (i.e., how often each exterior door was opened), and these were able to be used as a proxy for demand in the location model. In addition, response recommendations may change. As the pandemic progressed, small hand sanitizer pumps were also placed in each classroom. These additional resources affected the candidate locations for the pumps; the model shifted from being classroom-centric to locations near exterior doors. Iterative model development supported expanded access, i.e., doors and classrooms, as opposed to statically resulting in double-coverage.

**2.7 Creating an Entirely New Model**

**Definition**: This situation occurs when a new decision that needs to be made due to the surprise event needs an entirely new model to be done effectively.

**Overview and Examples**: This adaptation focuses on examining the environment after the surprise event, realizing that a new decision needs to be made that has never been made before, and creating a new model to address these decisions in collaboration between modelers and users.

In Summer 2020, Clemson University administrators were faced with an uncertain environment; students had already registered for courses for Fall 2020 based on the expectation that conditions would be "normal." In response to concerns about exposure to COVID-19, both students and faculty were given the option to be in class or be online, but political pressures meant that not all courses could be moved online. All faculty teaching in-person needed to engage in hybrid teaching. Rough estimates of classroom capacities indicated that not all currently registered students (that elected to be in-person) would be able to attend class in a socially distant manner as originally scheduled. At this point, the administration determined that "*something else had to change … a new "decision variable" was needed: students would not be able to attend every class meeting in person, but instead, students would only be allowed to attend a subset of class meetings in person.*"

In early summer 2020, the Facilities Department had not yet determined the actual classroom inventory for Fall 2020 courses, but the modelers started exploring how to model the new decisions:

> *"So I knew [that the facilities people] knew that we weren't going to be able to put everyone in the same rooms, because the capacity was going to be wonky. But we didn't know what the capacity was going to be, because it took a good while to actually map out the rooms and figure out how many rooms [were available]. ... Something we were doing on the side is [sic] we were using fall 2019 data that was stable to test and develop our models."*

The modeling team, composed of several OR faculty and graduate students, worked simultaneously on three modeling approaches: a graph theoretic approach, an integer programming approach and a heuristic approach:

> *"And so we had different models that we were trying. I kept trying to do a graph partitioning approach. I was just constantly trying to create partitions of students and find natural partitions and it was a disaster."*

After some iterations between the modeling team and the administration – the users (see Gore et al., 2022), a new integer program was developed and utilized in Fall 2020 and Spring 2021 to assign students to days in which they could attend courses in-person (a rotational attendance scheme):

> *"We were creating this top-down assignment scheme, and then we had a town hall. Somebody asked the question "how will I know what day my students will attend my class" and we already had an answer. We're going to give you that list of information."*

The simplified hierarchy of university course scheduling problems presented by Barnhart et al. (2022) does not include any sort of assignment of students to attendance days, illustrating the degree to which this presented an entirely new decision in response to COVID-19.

**Practical Considerations for Adaptation:** The situations in which an entirely new model may be needed echo those in which a toolkit-style adaptation may be appropriate: a new decision needs to be made and there is enough time in the build-up to the decision (or the decisions are adaptable after initially made) that a model can be implemented. However, in this style of adaptation the decision *does not* fit well enough within the context of a known OR model.

The importance of integrated development with users and stakeholders applies her as well. In this example, the modelers attempted to develop models based on several different frameworks but did not explicitly utilize an existing model. The modelers did not engage with the users and stakeholders in the model development; the results were primarily managed and interpreted by staff in the Registrar's Office, the Facilities Department and the Clemson Computing and Information Technology. In contrast with the hand sanitizer station project, the rotational attendance project did not lend itself to iterative improvement since it would not be easy to adapt the assignment of students to in-person class days; moreover, many stakeholders (faculty, students) were unavailable to provide feedback during the summer. This is an artifact of the type of decision that needed to be made, not the fact that a new model was needed.

## 3 Classification of OR-Informed Adaptations at Other Universities

We have presented a framework to classify the use of OR models to fundamental surprise events based upon interviews with modelers and users involved in Clemson's response to COVID-19. In this section, we apply our developed framework to classify how OR models were used in the response at other universities.

A sample of twelve papers, six on optimization models and six on simulation models, are presented in Table 3. They represent adaptation categories 3-5. We did not find papers that focus on adapting data or adding constraints, possibly reflecting a publication bias in terms of the requirements necessary to publish OR models in practice.

*Insert Table 3 approximately here (currently at the end of the paper).*

The optimization papers address classroom layout decisions (Murray, 2020; Greenberg et al., 2021), classroom assignments and modalities (Johnson and Wilson, 2021; Navabi-Shirazi et al., 2022), term planning and timetabling (Barnhart et al., 2022), and bus operations (Chen et al., 2021). One paper is classified in the "model switch" adaptation category (Chen et al., 2021). The bus-related decisions that the paper addresses, i.e., locations of hubs, stops, and routes, were also made pre-pandemic. However, COVID-19 changed the context surrounding these decisions; transit needed to be revisited with transmission risk in mind. In contrast, the classroom seating layout models (Murray, 2020; Greenberg et al., 2021) address new pandemic-induced challenges. Because they use facility location modeling tools, we classify them as "model toolkit" style adaptations. Similarly, the work of Johnson and Wilson (2021) utilize the classic assignment problem to analyze tradeoffs that arose during the pandemic to assign classes to classrooms. Class modality, classroom assignment, and the number of term decisions are also new and were addressed by Navabi-Shirazi et al. (2022) and Barnhart et al. (2022).

Among the simulation papers, Brook et al. (2021) address new decisions of campus asymptomatic screening and NPIs in the vein of existing model structures of COVID-19 branching processes (e.g., Hellewell et al., 2020). Frazier et al. (2022) evaluate policies to support return to campus using a SEIR modeling approach, i.e., a new set of decisions using existing modeling tools. New decisions related to community-risk of students returning home (Harper et al., 2021), locating testing stations (Saidani et al. 2021), other types of university operations (Ghaffazadegan, 2021), and classroom layouts (Islam et al., 2022) were evaluated with new models.

Many of the papers in Table 3 were used in an advisory manner to guide decisions leading up to the Fall 2020 and Spring 2021 semesters. This was especially true for one-time decisions for each semester (e.g., classroom layouts, course scheduling, testing station designs). Here, the models were used to support an adaptation by the university but are not necessarily designed to be adaptable in terms of their modeling characteristics. In some ways, it appears that fundamental surprise events are particularly well-suited to incorporate new models, or apply existing ones from the OR toolkit, into decision-making environments since the modeler-user (stakeholder) incentives are aligned to understand the impacts of decisions within a fundamentally new environment. The response to COVID-19 also allowed iterations between the modelers and users in many instances (for example, Johnson and Wilson, 2021; Navabi-Shirazi et al., 2022; Barnhart et al., 2022). However, the unique characteristics of COVID-19 may have driven the ability of such modeling efforts as well: there was a sufficiently long lead-time to the fall semester of 2020 from when the event occurred and there was a significant desire by many universities to move away from the initial response that occurred in the spring of 2020. Therefore, the time between the onset of the event and when new decisions need to be made also impacts the ability to build new models or incorporate existing models into decision-making.

## 4 Discussion

In this section, we discuss implications of our framework for understanding how modeling can support adaptation within organizations by connecting resilience engineering concepts to our framework. We recommend best practices for modelers as they seek to promote resilience to fundamental surprise events.

**4.1 Enabling Improvisation/Adaptation of the Modeler/User**

The framework described in this paper can be used as a roadmap for leveraging modeling capability to support adaptation in response to fundamental surprise. We view adaptation to these events from a cognitive systems perspective, centered around the domain user, i.e., the decision-maker (e.g., the Facilities Department), the modeler, the model itself, and the external dynamics observed in the world. Operationalizing the framework requires designing for interaction between each of these components. Successful use of the framework requires the modeler and the user to interface with each other in the context of external dynamics being monitored by each, jointly or separately. The modeler then matches the state of the world to the appropriate category of the modeling framework to determine the level of adaptation required (Table 2). The decision-maker as the 'end-user' of the model(s) provides input and feedback to the modeler based on their decision needs and to refine the real-world representation for the modeling. The modeler uses the feedback to iterate on their model or adaptation-category.

Figure 1 represents these notional relationships, including the feedback loop between the modeler and user, and the adaptive loops between each pair of entities. The modeler and user interact with the external dynamics of the world. This type of representation is like that illustrated by Eisenberg et al. (2019), which focuses on the use of resilience analytics to support adaptation to situational and fundamental surprise events. Our specific representation in Figure 1 focuses on situations of using models in response to fundamental surprise events where both the modeler and the user are monitoring the external dynamics of the event to understand how to adapt. This highlights one of the issues with solely relying on resilience analytics raised by Eisenberg et al. (2019) where data from the world (external dynamics) is directly fed into models without understanding context. We believe that Figure 1 is a specific application of the framework of Eisenberg et al. (2019) to how models can be used during the response to surprise events, supported by the evidence we gathered in creating our own classification framework.

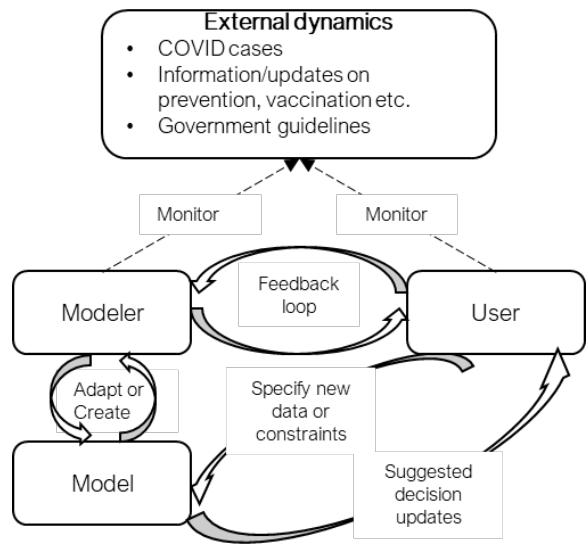

*Figure 1: Interactions between the user, modeler, model, and the external dynamics in the real world.*

From a cognitive engineering standpoint, the use of our modeling framework could be mapped to various human performance levels as defined in Rasmussen's Skills-Rules-Knowledge (SRK) framework (Rasmussen, 1983). Adaptations 1 and 2 are rule-based behaviors managed at the interface of the modeler, user, and the model. Adaptations 3-5 represent knowledge-based behavior as the real-world dynamics exceed the purview of any available representation. Adaptation at these levels entails identifying the "gap in knowledge" at an organizational level and then identifying, adapting, and/or creating a model to address it. In the future, a user-centered design approach could be used to support the interactions described above, including the cognitive needs of the modeler and decision-maker for rule-based and knowledge-based behaviors identified in the use of the modeling framework.

From an organizational adaptive capacity standpoint, the modeling framework provides the natural affordance to adapt decisions based on models. Having a framework in place, ready for use, helps mitigate the anchoring bias of fixating on a prior model with no clear criteria or path for revising and adapting (Tversky and Kahneman, 1974). The bias could result from not correctly identifying the shift in the external conditions or not having a roadmap available to respond to the shift with the appropriate adaptive response. The adaptive framework described in this paper helps enable readiness to respond and operationalize the concept of a system being poised to adapt (Woods, 2015).

### 4.2 On the Role of OR for Fundamental Surprise Events

Sharkey et al. (2021) discussed a research agenda for the network optimization community that would allow it to continue contributing to resilience, which can be extended to the broader OR community. It includes several areas that intersect with an understanding of how models are used during fundamental surprise event: *methods to handle 'unknown' uncertainty*, *models and/or methods that enable improvisation,* and *methods for model switching.* In our analysis, we found instances where improvisation and model switching occurred during the response to COVID-19 at Clemson University. Users improvised input parameters, such as capacity limits on classrooms, and input resources, such as altering the set of spaces that were considered classrooms, into decisions-making environments. Model switching occurred as hand sanitizer was deployed to cover all of campus at minimal cost, rather than cover as much of campus as possible under cost constraints. These observations may help to identify modeling guidelines that can be used to create approaches where the OR model can help impact resilience as graceful extensibility and adaptability since we have empirical evidence of how improvisation occurred across the modeler, user, and model dimensions.

An important takeaway from our analysis is that OR can play an important role in the response to fundamental surprise events by appropriately understanding the improvisations and/or adaptations that are being made in response to the event. This may require modelers to (quickly) collaborate with users of their modeling efforts to understand the type of support the users, or other stakeholders, are seeking in responding to the event and defining the scope of the model appropriately (Johnson and Wilson, 2021; Navabi-Shirazi et al., 2022; Barnhart et al., 2022; O'Brien et al., 2022). It also implies that OR should understand its own limitations in resilience, recognizing that resilience models for examining future events only handle situational surprise events – which is similar to the arguments made by Eisenberg et al. (2019). Similarly, it will be important for users to understand how the developed model is a simplification of the real-world decision context. Bridging from model-recommended decisions to implementation in practice should also embed user expertise and awareness not captured by the model.

OR can also use the knowledge gained during fundamental surprise events to help identify considerations that should be built into models to expand the set of models available to the field and its end users. Taking COVID-19 as a case study, safety considerations (such as social distancing) became critical in the operations of systems. For example, supply chains had been disrupted before COVID-19, but their recovery never had to consider the interactions between the workers at individual facilities. In general, there were not many models that incorporate safety considerations that were available, so this could be an area of future research to help support other pandemic-based surprise events.

### 4.3 Limitations of Our Approach

This study was limited by the fact that it focused only on the use of models at Clemson University during its response to COVID-19. While we do observe similar OR model uses at other universities during the pandemic, the empirical evidence upon which our framework is based is only from Clemson. There is sufficient evidence for each of the five classes of adaptations presented to be part of the framework; however, our framework may not be exhaustive since there could be other styles of modeling adaptations. Further, there is overlap between the authors of this paper and the modelers at Clemson University. We do not feel that this significantly impacts the classification framework or the empirical evidence supporting it. Overall, given the infrequency of fundamental surprise events, we would argue that our approaches to gather empirical evidence and then create the classification framework help to deepen our understanding of how OR models can be used in the response to these events.

## 5 Conclusions

This paper provided a framework of how OR models were used during university operations responding to the fundamental surprise event of the COVID-19 pandemic. Interviews with modelers, practitioners, and university stakeholders provided data on how models were adapted, switched, or created to address issues arising during this response. This data helped to provide a classification framework on of OR model use in response to fundamental surprise events. This classification framework helps to better understand how models can still impact events that cannot be anticipated in advance (and, therefore, cannot be modeled in advance). Further, it helps to provide a future research agenda on how to best enable the adaptations and improvisations by modelers and users to address fundamental surprise events.

There are several future research directions that could be pursued based on this work. First, OR modelers could use issues arising during COVID-19, such as safety considerations, to build out the modeling toolkit. This would shorten the time required to address similar issues should they arise in future fundamental surprise events since the pull from the model toolkit adaptation would replace the create a new model adaptation in these cases. Second, human factors research could be done on how to design interfaces between the user and the model to enable their ability to quickly implement the adapting data and adding donstraints adaptations. Third, research could be done to understand how OR model use throughout an organization impacts the ability of that organization to adapt to changing circumstances. Fourth, research could be done to understand how our framework could be applied across an event-continuum characterized by levels of variability and surprise. Events across this continuum would vary in terms of likelihood, severity, and scale, ranging from the familiar to fundamental surprise.

We hope this supports the OR community thinking about resilience and connecting to the broader resilience community. As well as vice versa, supporting resilience researchers in an appropriate use of modeling, where modeling has historically been seen as fragile, possibly with good reason.

**Acknowledgements**

This material is based upon work supported by the Clemson University Support for Early Exploration (CU SEED) grant, Tier 2, 2021-2022.

Willemain, T.R. (1994). Insights on modeling from a dozen experts. *Operations Research*, *42*(2): 213-222.

Willemain, Thomas R (1995). Model formulation: What experts think about and when. *Operations Research*, *43*(6): 916-932.

Woods, D.D. (2015). Four concepts for resilience and the implications for the future of resilience engineering, *Reliability Engineering & System Safety, 141*, 5-9.
**Appendix**

*Interview Protocol*

- Briefly describe your role and how it relates to the university's class scheduling.
- What were the main challenges faced at the start of the pandemic (around March 2020) in terms of decisions around class scheduling?
- Was there a contingency plan available for such a situation?
- What types of information were needed or sought regarding scheduling of classes? Were these available?

    <u>Probes</u>: Categories and sources of information

- How was this information shared?

    <u>Probes</u>: Media of communication – emails, dashboards, databases, messaging systems

- How did you know what information to look at and when?

    <u>Probes</u>: Government and external reports and guidelines (e.g. CDC, WHO, Federal, SC State)

- Did you consider students' and their families' perspectives during this process? How? What were some of the considerations that emerged?

> Probes: Safety; flexibility; accommodation (esp. for those coming from outside of Clemson)

- What were your main priorities in terms of scheduling classes?

  > Probes: Infection control; graduation times; student stress; resources for remote classes

- Did you use any data analytics or modeling in the scheduling? Please describe this process.

- How did you decide what might be an 'optimum' number of students who can be on campus? Was there a priority, for instance, to accommodate the maximum no. of students who wanted to (or could possibly want to) be on campus?

  > Probes: no. of classrooms or spaces; classroom size; no. of students who wanted to be on campus vs. remote; other

- How did variability and uncertainty influence these methods?

  > Probes: changes in advisories/guidelines; infection rates; student attendance; other

- Was there any simulation or testing involved?

- Once classes resumed (remotely or hybrid), was there a way to monitor how things were going?

  > Probes: attendance; infection rates; testing protocols; compliance; other

- Did plans/strategies have to be adjusted based on actual dynamics of classes and student-presence on campus? What were some of these changes?

- Looking back, what types of information/data have been most important or useful in this process? What additional data or information would have been useful?

- How was information and decisions communicated across institutional layers – President and Provost's offices; college-level leadership; departmental leadership; other administrative offices?

    Probes: Emails, townhalls, dashboards, meetings/taskforce

- How were decisions and implementation actions coordinated across various bodies? Who were the facilitators?

| Style | Reference | University | Problem | Model/Approach | Implemented? | Type of Adaptation | | | | |
|---|---|---|---|---|---|---|---|---|---|---|
| | | | | | | Adapt Data | Add Constraints | Model Switch | Model Toolkit | New Model |
| Optimization | Murray, 2020 | University of California, Santa Barbara | Classroom physical distancing (capacity and layout) | Location modeling | Unknown | | | | x | |
| | Chen et al., 2021 | Michigan | Bus operations to reduce transmission (locations, routes, schedules) | MIP; simulation | Yes | | | x | | |
| | Greenberg et al. 2021 | Cornell | Classroom seating layout | IP and computer vision | Unknown | | | | x | |
| | Johnson and Wilson, 2021 | Oklahoma State University | Assigning hybrid classes to classrooms while considering rotations of students | IP | Yes | | | | x | |
| | Navabi-Shirazi et al., 2022 | Georgia Tech | Decide class modality; assign classrooms for in-person classes | MIP | Advisory | | | | | x |
| | Barnhart et al., 2022 | MIT | Term planning (which terms to offer courses in) and timetabling (scheduling the courses) | IP | Advisory and implemented | | | | | x |
| Simulation | Harper et al., 2021 | Cardiff | Estimate infections if students return home | Monte Carlo simulation | Advisory | | | | | x |
| | Saidani, et al., 2021 | University of Illinois Urbana-Champaign | Testing stations and operators (how many and where) | Discrete event simulation | Yes | | | | | x |
| | Ghaffarzadegan, 2021 | Virginia Tech | University operations (testing, mask use, communication, remote work) | Simulation | Unknown | | | | | x |
| | Brook et al., 2021 | Berkeley | Asymptomatic surveillance testing and NPIs | Stochastic branching process | Advisory | | | | x | |
| | Frazier et al, 2022 | Cornell | Student return to campus (spread, testing, and contract tracing) | Compartmental simulation | Advisory | | | | x | |
| | Islam et al., 2022 | Arizona | Pedestrian movement in classroom-style indoor areas (how to allocate seats, layout, entrance/exits) | ABM | Unknown | | | | | x |

*Table 3: Classification of a sampling of OR papers for university response to COVID-19 within our framework*